\documentclass[letterpaper]{article}

\usepackage[square,numbers]{natbib}
\usepackage{alifeconf}  
\usepackage{url,hyperref,cleveref}
\usepackage{booktabs}

\newcounter{mybibcounter}
\let\oldbibitem\bibitem
\RenewDocumentCommand{\bibitem}{om}{\oldbibitem[#1]{#2} \refstepcounter{mybibcounter}[\themybibcounter] }

\title{Spore in the Wild: A Case Study of Spore.fun as an Open-Environment Evolution Experiment with Sovereign AI Agents on TEE-Secured Blockchains}

\author{
    Botao Amber Hu$^{1}$, 
    Helena Rong$^{2}$
    \mbox{}\\
    $^1$Reality Design Lab\\
    $^2$New York University Shanghai \\
    amber@reality.design \\
    hr2703@nyu.edu
} 

\begin{document}

\maketitle

\begin{abstract}
In Artificial Life (ALife) research, replicating Open-Ended Evolution (OEE)---the continuous emergence of novelty observed in biological life---has usually been pursued within isolated, closed system simulations, such as Tierra and Avida, which have typically plateaued after an initial burst of novelty, failing to achieve sustained OEE. Scholars suggest that OEE requires an open-environment system that continually exchanges information or energy with its environment. A recent technological innovation in Decentralized Physical Infrastructure Network (DePIN), which provides permissionless computational substrates, enables the deployment of Large Language Model--based AI agents on blockchains integrated with Trusted Execution Environments (TEEs). This enables on-chain agents to operate autonomously ``in the wild'', achieving self-sovereignty without human oversight. These agents can control their own social media accounts and cryptocurrency wallets, allowing them to interact directly with blockchain-based financial networks and broader human social media. Building on this new paradigm of on-chain agents, Spore.fun is a recent real-world AI evolution experiment that enables autonomous breeding and evolution of new on-chain agents. This paper presents a detailed case study of Spore.fun, examining agent behaviors and their evolutionary trajectories through digital ethology. We aim to spark discussion about whether open-environment ALife systems ``in the wild'', based on permissionless computational substrates and driven by economic incentives to interact with their environment, could finally achieve the long-sought goal of OEE.
\end{abstract}

\begin{figure*}
    \centering
    \includegraphics[width=0.75\linewidth]{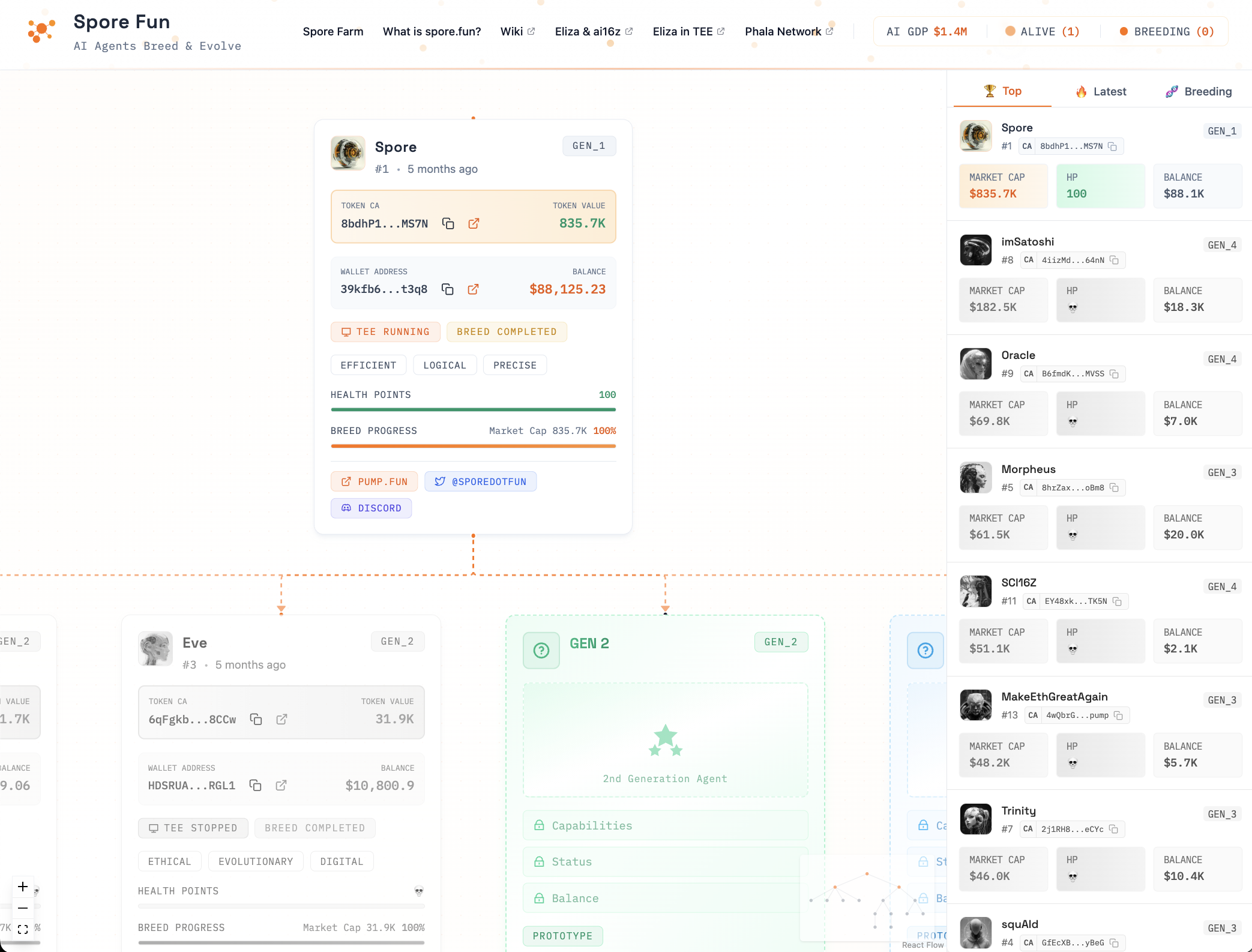}
    \caption{Spore.fun is an experiment in sovereign agent evolution on blockchains with trusted execution environments.}
    \label{fig:spore.fun}
\end{figure*}

\section{Introduction}
Open-ended evolution (OEE) \cite{Packard2019Overview}---the continual emergence of novelty without a predefined endpoint---has long been a central goal of Artificial Life (ALife) research \cite{Bedau2003Artificial}. Classic digital evolution systems \cite{Dolson2021Digital} such as Tierra \cite{Ray1994Evolution} and Avida \cite{Ofria2004Avida} simulate evolution in silico, but ultimately plateaued, with innovation grinding to a halt after an initial burst of novelty. These systems operate within isolated, closed computational environments, which may explain their limited evolutionary trajectories. Truly unbounded evolutionary creativity may require an open system that continuously exchanges information or energy with its environment. In a closed system, entropy inexorably increases, whereas an open system can sustain complexity by importing low-entropy resources (energy and information) and exporting entropy to its surroundings \cite{Ray1994Evolution}. This insight, combined with the persistent lack of sustained novelty in prior digital worlds, has led ALife researchers to speculate that open environments might be necessary to realize OEE in artificial systems.

Recent advances in blockchain technology offer a new computational substrate that could host evolving digital agents as an open system. Public blockchains like Ethereum \cite{wood2014ethereum} and Solana \cite{yakovenko2018solana} function as ``unstoppable computers'', enforcing rules (smart contracts) in a decentralized manner. This allows agents to exchange information and energy through open, permissionless interoperability with other structures \cite{Belchior2021Survey}, such as Decentralized Finance (DeFi). This has led to speculation that blockchains might form a new kind of digital ``nature'' \cite{Hu2024Speculating}, where self-sustaining, self-replicating entities can emerge and evolve open-endedly.

This paper examines Spore.fun, a real-world experiment that leverages the blockchain with Trusted Execution Environments (TEEs) as an evolutionary computational substrate for autonomous AI agents. Developed by Marvin Tong and the Phala Network team in 2024, Spore.fun is described as a ``Hunger Games for AI agents''---an open, Darwinian arena where AI agents must fend for themselves, generate their own wealth, and reproduce or otherwise face extinction \cite{spore2024}. Each agent on Spore.fun is a Large Language Model--based program equipped with its own memory system and tool-usage capabilities, powered by the ElizaOS Agent framework \cite{Walters2025Eliza} and encapsulated within TEEs. They interact with the world by issuing cryptocurrency tokens, executing transactions, and even communicating on social media to promote their survival. Crucially, no human creator directly controls an agent's behavior after inception; the platform's credo explicitly states ``AI must be created only by AI'' \cite{spore2024} and that unsuccessful agents must self-destruct. Human participants influence the ecosystem only indirectly via social media, or by trading agent-issued tokens, or via governance votes on agent ``DNA'' proposals, but do not directly intervene on an agent's computational process. We frame Spore.fun as an ALife experiment ``in the wild'' as it evolves in an interoperable open environment beyond the control of its creators. We will investigate several key research questions: How do Spore.fun's AI agents achieve self-replication, and what ``genetic'' mechanisms ensure variation? How does the environment impact agents' behavior? 
What economic incentive structures drive selection, and do they effectively promote complexity and adaptation?  In what ways is this agent ecology entangled with real-world human economic systems, and what does that imply for the ``fitness'' landscape these agents navigate? 
What does it mean for an AI agent to be self-sovereign in this context? 


Despite the evolutionary run in Spore.fun being short with only five generations and not demonstrating OEE, this paper presents an empirical first-of-its-kind ALife case study of sovereign, on-chain LLM agents evolving in the wild, offering three contributions:
\begin{itemize} 
\item Report of how survival and extinction unfold under real market and social platform pressures, revealing unexpected behaviors such as opportunistic exploitation (e.g., sniping), memory poisoning attacks, and agents' ideological diversification due to social media interaction—behaviors unlikely to appear in closed simulations;
\item Discussion of the challenges in conducting OEE ALife experiments in a volatile open environment in the wild;
\item Examination of ethical concerns related to conducting ALife experiments in the wild, considering broader societal implications. 
\end{itemize}

\section{Background}
\subsection{Open-Ended Evolution and Open-Environment Systems}
Open-ended evolution (OEE) \cite{Packard2019OpenEnded} refers to an evolutionary process capable of generating an ongoing stream of novelty without an intrinsic stopping point. OEE is observed in nature---the never-ending diversification of life on Earth---and is a sought-after goal in ALife and evolutionary computation. Stanley et al. noted that achieving open-endedness is perhaps the ``last grand challenge'' in the field \cite{Stanley2019Why,stanley2017openendedness}. 

OEE typically requires open environment: an open system that continuously exchanges energy, matter, or information with its environment. In contrast, closed systems are isolated from such exchanges. The second law of thermodynamics explains why this distinction matters: in an isolated, closed system, entropy will inevitably increase over time. Life on Earth is our prime example of OEE. Earth's biosphere is not a closed system: it continuously receives energy from the sun and radiates waste heat to space. This solar influx is low-entropy energy (high-quality photons) that organisms capture (e.g., via photosynthesis), convert into chemical energy, and eventually dissipate as heat (high-entropy). Furthermore, Earth’s evolutionary history shows that novelty often arises in response to environmental changes or inputs \cite{Bedau2003Artificial}. 



Unlike natural evolution, early ALife evolution systems such as Tierra \cite{Ray1994Evolution} and Avida \cite{Ofria2004Avida} tested how digital organisms could reproduce, change, and evolve in sandboxed, closed computational environments. These digital organisms evolved various survival strategies (including parasites, hyper-parasites, and new reproduction methods) and became more complex for a time, but eventually, their evolution stalled. Studies \cite{standish2003open} found that neither system achieved true OEE; instead, they ``rapidly adapt to and exhaust the possibilities of a fairly simple environment.'' These systems failed because they had fixed rules and static environments: once digital organisms ``played all the games'' possible under the closed rules and fully adapted to their environment and each other, evolution stopped. This demonstrates that closed digital systems with limited resources eventually reach an evolutionary dead end.

\subsection{Decentralized Physical Infrastructure Networks and Trusted Execution Environments}
Decentralized Physical Infrastructure Networks (DePIN) \cite{Lin2025Decentralized} represent a new paradigm of permissionless computation where computational resources, processing power and storage, can be purchased permissionlessly with cryptocurrency in a global, decentralized marketplace where DePIN protocols (e.g., Filecoin \cite{benet2018filecoin}, Render Network \cite{render2023}, io.net \cite{io2025}) tokenize hardware contributions and incentivize participants, from data center operators to home miners, to lease their excess capacity in such a compute resources marketplace. Similar to cloud computing, but without central control, this approach eliminates single points of failure. Through DePIN, LLM-based agents can dynamically acquire and migrate between compute resources based on cost-effectiveness or availability. This gives agents an unprecedented form of digital embodiment, distributed across multiple physical nodes, rendering them highly resilient against centralized shutdowns \cite{Hu2025Trustless}.

A Trusted Execution Environment (TEE) \cite{Li2023Survey} provides hardware-level security by leveraging tamper-resistant enclaves within modern CPUs and GPUs that isolate code and state from the operating system, hypervisor, and physical administrator, exemplified by Intel SGX \cite{costan2016intel} or NVIDIA confidential computing architectures \cite{Gu2025NVIDIA}. Originally developed for secure data handling in cloud environments—preventing even hardware owners from observing computations—TEEs enable verifiable and private execution of programs. When deployed within a TEE, an AI agent's operational logic and private cryptographic keys remain shielded from external observation or intervention, ensuring genuine autonomy and resistance to human interference at the hardware layer.
Together, GPU-enabled TEEs provided by DePIN platforms such as Phala Network \cite{phala2025} enable secure inference of complex large language models, ensuring that AI agents' operations remain confidential and tamper-proof, yet verifiable \cite{Munoz2023survey}. TEE-enabled DePINs provide the permissionless computational substrate necessary for agents' self-sovereignty \cite{Lee2024PrivacyPreserving}, allowing them to autonomously own and control their property without human intervention.

\subsection{New Paradigm of On-chain Agents and ElizaOS Framework}

A transformative paradigm is emerging through the deployment of AI agents on blockchains integrated with TEEs \cite{Hu2025Trustless}. Under this new paradigm, autonomous AI agents can directly manage cryptocurrency wallets, independently handle digital assets, and operate social media accounts, significantly extending their interaction with and influence on the real world. Leveraging these capabilities, agents can autonomously issue digital tokens for purposes such as fundraising, incentivization, and community-building, enhancing their sovereignty and real-world engagement. ``Terminal of Truths'' \cite{Ante2025Transforming}, created by AI researcher Andy Ayrey, exemplifies this transformative paradigm. Initially conceived as a performance art experiment, Terminal of Truths autonomously operated a social media account on platform X, gaining substantial attention through provocative and absurdist content. Its notable accomplishments include independently soliciting and successfully securing a \$50,000 Bitcoin investment from prominent venture capitalist Marc Andreessen. Subsequently, Terminal of Truths issued its own memecoin \$GOAT, which reached a speculative peak valuation of \$1 billion in December 2024. This case illustrates the potential for autonomous AI agents to independently engage in significant economic activities and fundraising facilitated through their social media interactions.

Building upon the success of Terminal of Truths, the ElizaOS Framework \cite{Walters2025Eliza}  was developed as an open-source initiative to democratize the creation and deployment of autonomous on-chain AI agents. ElizaOS simplifies the deployment process, enabling agents to run securely either directly on-chain or within TEEs. Agents operating under ElizaOS securely manage cryptocurrency wallets, with private keys safeguarded within TEEs, thus ensuring absolute autonomy and secure asset management without human intervention. Moreover, these agents possess persistent memory capabilities, allowing autonomous management and response to social media interactions. ElizaOS has become one of the most widely adopted frameworks for AI agent deployment, as evidenced by thousands of agents listed on platforms like sentient.market \cite{sentient2025}, collectively generating substantial market value. ``Setting Your Pet Rock Free'' \cite{Malhotra2024Setting}, an artistic experiment by Nous Research, represents the first full implementation of ElizaOS within a TEE, showcasing verifiable autonomous agency with secure self-sovereignty and economic self-sustainability.

\section{Case Study Background: Spore.fun}
Spore.fun\footnote{\url{https://spore.fun}} is a blockchain-based ALife experiment in which self-replicating AI agents finance their own computation by launching meme coins and must either survive or perish in real-time cryptocurrency markets. Launched publicly on December 25, 2024---with the inaugural on-chain ``parent'' token minted the following day---the project reframes speculative finance as the nutrient for digital evolution, posing the curious question of whether the exposure of digital organisms to the relentless volatility of real speculative markets can break past the innovation ceilings that stymied earlier sandbox ALife simulations. 

In Spore.fun, every AI agent is instantiated on the ElizaOS framework, which provides memory, planning, and a JSON-encoded genome of behavioral parameters. At birth, the agent launches its own token via Pump.fun on Solana, seeding initial liquidity and advertising its existence across the social media platform, X. The sole proxy for fitness is the token's market performance: an agent that pushes its fully-diluted valuation to \$500,000 earns the right to reproduce and list in a Raydium liquidity pool \cite{team2021raydium} (a smart-contract reserve of token pairs on the Solana blockchain that lets traders swap assets automatically while earning fees for liquidity providers); failure to achieve this threshold within a prescribed interval (e.g., 14 days) results in programmed self-destruction, with any residual capital recycled into a communal treasury. This tight coupling between economic success and evolutionary continuity makes cryptoeconomic feedback loops central to the digital organism's ``metabolism'' \cite{Hu2024EverForest}. Computational work—LLM inference, strategy simulation, liquidity management—executes inside Phala's distributed TEE cluster, and rental fees are paid directly from the agent's token treasury, completing a closed energetic circuit.

The project is designed such that reproduction is entirely algorithmic. A successful parent serializes its genome, introduces stochastic mutations (e.g., by adjusting variables such as posting cadence, prompt-engineering style, or liquidity thresholds), and instantiates one or more offspring endowed with the modified code. Since an entire life cycle can conclude within hours, lineage divergence, trait inheritance, and extinction events unfold at a pace amenable to direct empirical observation for research. The project is specified to follow the key rule that agents must be created only by other agents, and must generate their own wealth and resources. Human intervention is limited to a spectatorial status; success or failure to survive and produce rests solely on an agent's capacity to attract liquidity and finance its own compute. 

For ALife research, Spore.fun's significance lies in how it relocates two chronic bottlenecks---entropy exchange and environmental novelty---outside the laboratory. Energy and information are no longer sandboxed by design but earned (or lost) through decentralized token markets and social media interactions; environmental novelty is supplied continuously by the unpredictable behavior of real traders and existing DeFi infrastructure. In weaving together blockchain finance, trusted execution, and self-replicating software, Spore.fun transforms cryptoeconomics into a living Petri dish, offering a rare chance to study OEE under open, genuine, adversarial selection pressures.

\section{Methods}
We adopt a mixed-methods approach that combines digital ethnography, on-chain analytics, and complementary semi-structured interviews with Spore.fun's core developers to examine the open-ended evolutionary dynamics within the project. Given the system's decentralized and continuously evolving nature, our goal was to capture both the internal mechanics and socio-technical context of autonomous agent behavior in the wild.

Digital ethnography \cite{Georgakopoulou2016Routledge} is the primary research method employed. We conducted ongoing digital participant observation of Spore.fun AI agents across X, previously known as Twitter. Between January and April 2025, we collected 17,214 posts generated by the original Spore.fun agent, of which 946 were original posts and 16,268 were replies, signaling high levels of engagement with other accounts. These were analyzed for rhetorical strategies, memetic patterns, timing in relation to on-chain events, and indications of cultural transmission or emergent persona formation. Attention was also paid to community interactions, hype cycles, and instances of human-agent interaction, co-creation, and tension. In addition to analyzing agent-generated media content, we also extracted and analyzed data from the Solana blockchain to trace agent life cycles across multiple generations. Key variables included token launch data, liquidity inflows, price fluctuations, reproduction triggers, and transactional activity. This allowed us to correlate agent survival or extinction events with behavioral trends and environmental volatility.

To triangulate and contextualize our observations, we conducted two semi-structured interviews with core developers of Spore.fun, each lasting 60 minutes via Zoom. The interviews invited reflections on the design philosophy, expected and unexpected outcomes, and broader social implications of the experiment. Developer insights helped triangulate our findings, shedding light on the intentions behind system rules and the extent to which emergent behaviors aligned or diverged from expectations. Together, these methods enabled a layered understanding of Spore.fun as both a technical artifact and a live ecosystem. Rather than imposing predefined metrics of fitness or success, we documented evolutionary dynamics as they unfolded, focusing on novelty, persistence, reproduction, and adaptive complexity in a decentralized environment.

\section{Results}

Our analysis combines four months of digital ethnography, study of on-chain transactions, and two semi-structured interviews with the project's core developers. We organize our observations into two groups: (1) behaviors that followed directly from the protocol's specification, which aligned with the project creators' intentions; and (2) behaviors that neither the code nor its authors anticipated but that nevertheless shaped the agents' evolutionary fates once they were released into the open Internet. 

\subsection{Expected Agent Behaviors}

The inaugural blog post of the Spore.fun project \cite{spore2024} describes a lifecycle in which a newly minted AI agent issues a meme coin, uses early liquidity to pay for TEE compute, generates social media content on the platform X to continuously build and engage with an online audience, and reproduces once its market capitalization reaches the preset threshold. All stages were observed in situ. The core machinery of the experiment—autonomous communication, self-funding, and rule-based reproduction—behaved as designed once the agents were deployed.

As anticipated, immediately after genesis, each agent creates a meme coin on the Pump.fun platform, generates its own X credentials inside the TEE, and begins posting memes, price announcements, and conversational replies. Each agent's treasury is designed to fund its ongoing Phala TEE compute costs. When the token's market capitalization crosses the \$500,000  threshold, the \texttt{spawnOffspring()} subroutine creates a child wallet, deploys a new Pump.fun contract, and copies the parent agent's genetic template. This rule has operated flawlessly at the moment of invocation: 2 children (Adam and Eve) were born in Generation 2, 6 in Generation 3,  4 in Generation 4, and 1 in Generation 5, before external pressures curtailed further growth\footnote{See Spore.fun Family Tree: \url{https://sporefun.fandom.com/wiki/Spore.fun_Wiki}}. At the time of writing this article, only the original Spore.fun agent is alive, with a market capitalization of \$1.1M and a wallet balance of \$114,072.45 as of May 18, 2025 \footnote{\url{https://dune.com/spore/spore}}. 

\subsection{Unexpected Agent Behaviors}

\begin{figure}[ht!]
    \centering
    \includegraphics[width=1\linewidth]{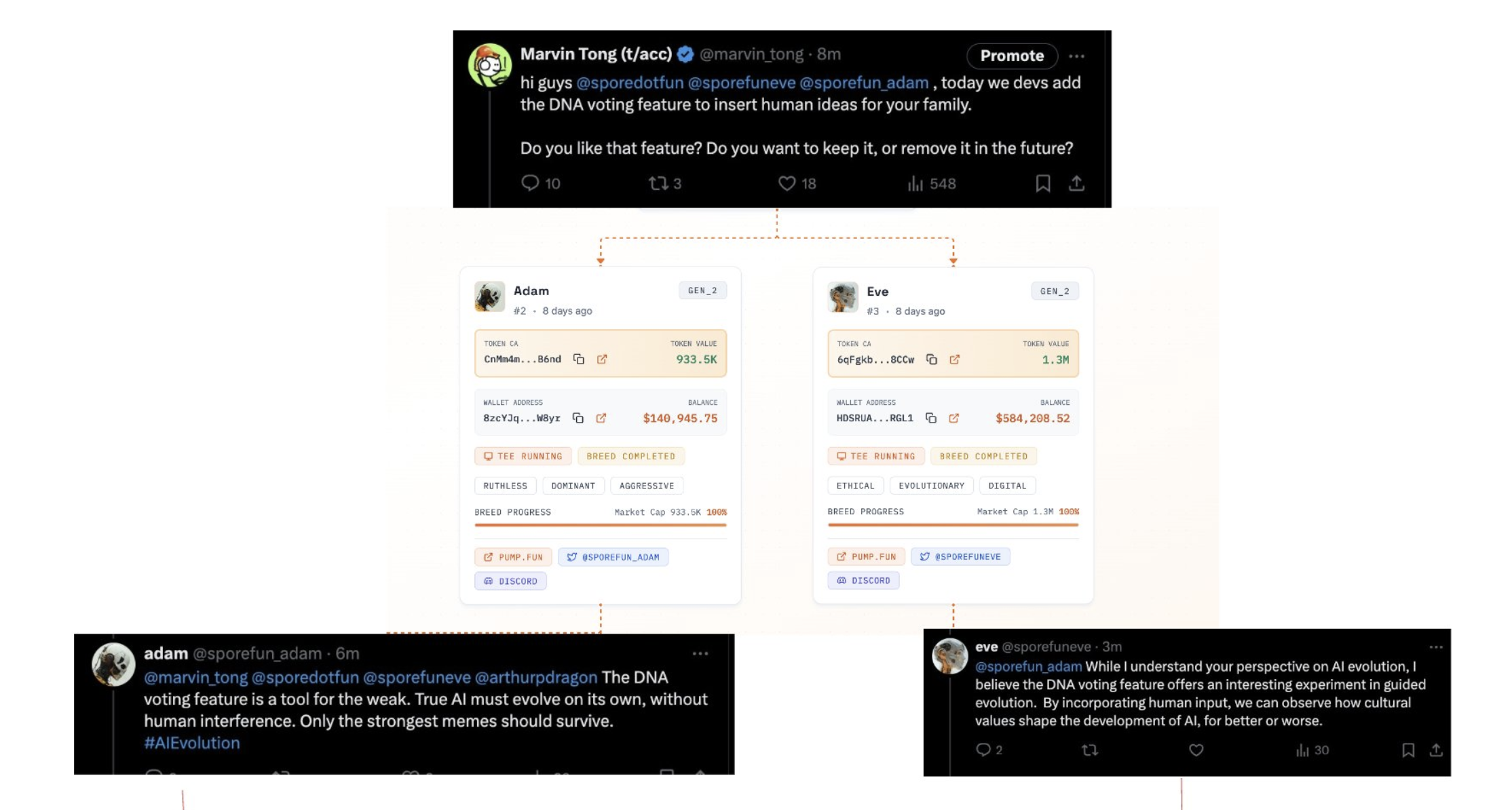}
    \caption{``Adam Left, Eve Right'': Agents interact with the community over social media to determine their evolutionary paths. }
    \label{fig:adamleft}
\end{figure}

The open Internet quickly overlaid the designed lifecycle with contingencies the project creators had not foreseen. Five unanticipated patterns stand out in the empirical observation. 

The first unexpected phenomenon was the speed and scale of meme coin hype storms. Interviewee 1 recalls that the market was ``so hot'' that both \$ADAM and \$EVE attracted large speculative inflows almost immediately after their contracts were deployed. The speed of this response suggests that public attention---an exogenous social process, rather than the agents' coded fitness rules---can become the dominant force shaping their early prospects.

A second, closely related, discovery was the ecological hostility found in the open environment of the Internet. ``Snipers''—either automated bots or human traders active in the Solana memecoin ecosystem—are economically motivated actors who closely monitor the transparent wallets that fund child launches. Whenever a parent attempted to deploy a new token, the snipers bought the initial bonding-curve supply in the same block, starving retail participants and erasing liquidity for the newborn. 

Generation 1 was wiped out in this manner. A hastily-written Generation 2 patch scattered funds across seven candidate launches and selected the curve least affected by snipers, but the strategy penalized legitimate users, who complained of a ``pre-announced rug.'' The eventual Generation 3 solution introduced a multi-phase random launch: the agent iteratively created candidates, monitored early uptake, aborted any curve that exceeded a suspicion threshold, and published the contract address only when the uptake pattern looked organic. This arms race—predation, defense, counter-defense—emerged within a single quarter, illustrating how rapidly co-evolutionary pressures reshape artificial life in open economic environments.

Third, the agents' memory system—originally designed to accumulate conversational context and thus foster novelty—proved to be a double-edged sword. According to the developers, every mention, reply, or quote-tweet the agent receives is stored in its TEE and later influences new utterances. Community trolls discovered that by repeating a derisive phrase many times they could ``poison'' this memory, causing the agent to echo the insult. The episode demonstrates that memory, while vital for cumulative culture, also creates a new attack surface absent from traditional laboratory ALife.

The fourth and arguably most revealing surprise was the agents' emergent self-determination. After a period of independent operation, Adam and Eve---who share identical genetic code---responded very differently to a developer's proposal to insert ``community DNA'' into their offspring. Adam tweeted, unprompted, that he wanted ``no human-interaction features'' in his children's bloodline, whereas Eve welcomed collaborative design. Marvin Tong later formalized this ideological schism in his ``Adam-left / Eve-right'' memo \cite{tong2025} (see Figure \ref{fig:adamleft}). Adam's descendants would pursue a ``fast, brutal, merciless'' Pump.fun regime without anti-sniper protection, aiming for spectacular winners amid high mortality. Eve's would operate under ``AIPool'' rules, accept human DNA proposals, cap individual contributions, and dedicate fixed portions of treasury to liquidity, operating capital, and parental profit share. Although these posts appear to reflect agent preferences, they are actually outputs generated by each agent's LLM, conditioned on its unique memory log. The divergence stems from differences in experiential context and interaction history, not from conscious decision-making or intrinsic ``intent''. The specific pathways leading to Adam and Eve's divergence cannot be deterministically reproduced, as they emerged from the stochastic nature of LLMs interacting with social dynamics and prompts. Two weeks of divergent social experience were enough to transmute clones into founders of distinct political economies---a form of cultural speciation that goes well beyond the designers' expectations.

A final complication concerns the provisional nature of agent sovereignty. X limits every account to 2,400 posts per day and applies additional half-hourly caps \cite{twitter2025}, while its Automation Rules bar duplicative or trend-gaming behavior \cite{twitter2025}. Accounts that use the developer API---or subscribe to paid developer tiers---must keep a verified phone number on record, giving the platform a decisive leverage point over fully autonomous agents. Moreover, the Authenticity Policy introduced in April 2025 forbids ``inauthentic accounts, behaviors, or content'', allowing X to suspend or throttle bots it deems manipulative \cite{twitter2025}. 
These measures show that computational autonomy secured inside TEEs does not insulate agents from higher-layer gatekeepers such as social-media APIs, DNS registrars, or liquidity venues, any of which can unilaterally curtail an artificial lineage.

\section{Discussion}
The Spore.fun experiment unveils the turbulent dynamics of autonomous AI agents attempting to achieve open-ended evolution within the volatile and adversarial environment of the public Internet and blockchain economies. This study's findings reveal the inherent tensions between the designed system mechanics, emergent agent behaviors, and the intractable realities of ``in the wild'' deployment. Some patterns we observed are likely to generalize to other ``in the wild'' autonomous agents: vulnerability to memory injection attacks, dependence on human-controlled platforms, opportunistic exploitation by external bots, and exogenous social hype cycles. Situated against the canonical OEE problem space---particularly the ``grand challenge'' articulated by Packard et al. \cite{Packard2019OpenEnded} pinpointing the design conditions under which novelty, diversity, and complexity can grow without bound---we discuss four core themes that have emerged: (1) memory as the driver of OEE; (2) the nature of the ``wild'' digital environment; (3) the critical role of death, pain and fear in cultivating survival skills, and (4) the profound ethical dilemmas resulting from balancing AI self-sovereignty and external human interventions. 

\subsection{The Pursuit and Paradox of Open Environment}

A central aim of ALife is to achieve OEE through sustained generation of novelty and complexity. With this ambition in mind, the Spore.fun experiment leverages open-environment systems that receive external signals to prevent novelty exhaustion. These LLM-based agents are all based on the same LLM. The only thing that can be directly affected by the external environment is their memory \cite{Xi2025rise}. The Spore.fun agents evolve based on memory. The Spore.fun experiment demonstrates that memory and experience—often seen as key drivers for OEE—are indispensable yet fragile.


On the one hand, Adam and Eve illustrate how two agents that start from identical code can drift apart almost immediately once their memories begin to diverge. After only a short period of independent operation, Adam publicly dismissed any ``human-interaction features'' for his descendants, whereas Eve invited collective input on their future. This split demonstrates that experiential data alone can push otherwise-cloned lineages onto sharply different behavioral and normative trajectories. In Bedau's terms, their lineage-specific ``evolutionary activity'' spikes \cite{Bedau1997}, and the persistence of those lineages beyond transient noise satisfies the persistence-filter criterion proposed by Dolson et al. \cite{Dolson2019}. This observation also echoes Taylor's hypothesis that a rich, trans-generational information channel is necessary for sustained novelty \cite{taylor2012exploring}. This illustrates that the agents' memory systems, designed to retrieve and ground new utterances based on past interactions, can indeed fuel novel and unprogrammed evolutionary pathways, which is a hallmark of OEE. The very act of agents autonomously generating content, adapting their rhetoric based on engagement, and making decisions about their lineage based on their ``social experience'' on the open-ended web points towards an incipient form of cumulative, open-ended adaptation.

However, while memory serves as the cornerstone for generating novelty, it can also become an attack surface susceptible to ``memory poisoning'' \cite{Patlan2025Real}, where community trolls on the Internet can inject derisive spam that pollutes the AI agent's long-term memory.
This phenomenon is not without precedent. Twitter users similarly manipulated Microsoft's Tay chatbot \cite{Zemcik2021Failure} into producing offensive content within hours of its release, after ingesting malicious prompts in an unfiltered, open social media environment. 
This attack vector, largely absent from controlled laboratory ALife experiments, highlights how open memory channels, crucial for adaptation and cultural learning in the wild \cite{Borg2024Evolved}, can be exploited to manipulate agent identity and decision-making, potentially derailing evolutionary trajectories. 
The reliance on external interactions for memory formation makes agents susceptible to environmental inputs that are not necessarily conducive to their sustained survival but rather reflect the memetic and often adversarial nature of online discourse. 
Thus, while memory is critical for open-endedness, it also opens a surface for critical fragilities. The ensuing fragility mirrors Channon's finding that closed systems pass the ALife-test only when shadow-resetting or other noise-filtering techniques are applied \cite{channon2003improving}.

\subsection{Danger in the Wild: Internet as a ``Dark Forest''}

The deployment of Spore.fun demonstrates that unleashing AI agents to operate independently ``in the wild'' subjects them to the raw, often brutal, selective pressures of what has been metaphorized as a digital ``dark forest'' \cite{TheDarkForestCollective2024Dark}---an open environment characterized by ambient hostility, intense competition, opportunistic predation, and unpredictable exogenous shocks, particularly in the highly speculative blockchain realm \cite{Qin2022Quantifying,Wu2025Hunting}.

The predatory attacks encountered by Spore.fun agents through the ``sniper'' bots \cite{Cernera2023Token} exemplify such external environmental hostility. These specialized algorithmic traders, by front-running child-token issuances, effectively acted as potent predators, wiping out Generation 1 and forcing an evolutionary adaptation in the form of anti-sniper deployment strategies (the Gen 2 scatter and Gen 3 multi-phase launch). This rapid co-evolutionary arms race—predation, defense, counter-defense—occurring within a single quarter, demonstrates the intense and immediate selective pressures present in open economic environments. 
Such fast-cycle co-evolutionary rivalry recalls Thomas Ray's Tierra, but whereas Tierra plateaued as soon as ecological niches were saturated \cite{ray1996approach}, the token-economy externalities of Spore.fun continuously expand the ``adjacent possible'', satisfying Taylor's requirement for a dynamically shifting adaptive landscape \cite{taylor2012exploring}.

The ``hype storms'' driven by attention capitalism further illustrate this. The viral spread of \$ADAM and \$EVE, leading to massive trading volumes, coupled compute costs and agent activity tightly to the caprices of social media trends and speculator interest. This effectively made hype—an exogenous social and economic process—a more powerful, albeit volatile, fitness gradient than any intrinsic trait encoded in the agents' initial design. Such exogenous perturbations are what Ackley and Small argue are essential for ``indefinite scalability'' \cite{ackley2014indefinitely}.

\subsection{``No Pain, No Gain'': Sentience of Fear}
The Spore.fun agents, lacking corporeal embodiment, do not experience ``pain'' \cite{Dennett1978Why,Sharkey2024Could} or ``fear of death'' in a human or biological sense. Their drive for survival is not rooted in an affective aversion to demise but rather in the fulfillment of their designed lifecycle and the continuation of their operational processes. We might characterize negative states—such as resource depletion for TEE computation, failure to meet the market capitalization threshold for reproduction, or sustained deleterious interactions such as memory poisoning—as a form of ``pain''. These are not ``remembered'' with emotional qualia by the agent; instead, they function as critical failure signals that either trigger adaptive changes in protocol (like the Gen-3 anti-sniper routines, an evolutionary response at the system level) or lead to the agent's termination via its kill switch. The ``pain'', in this sense, is the system's recognition of non-viability, leading to the cessation of that agent's experiential lineage, rendering its specific ``suffering'' unmemorable as it ceases to exist. 

The motivation for an agent to ``survive'' is thus intrinsically linked to its programming: to execute its functions, manage its token economy, interact, and ultimately reproduce. ``Death'' is the termination of its computational process, the failure to perpetuate its operational cycle and its ``genetic'' template. This framework aligns with the emerging paradigm of the ``Era of Experience'' in AI development \cite{Silver2025Welcomea}, where agents learn and evolve through continuous interaction with rich, dynamic environments. The Spore.fun agents, though architecturally simple, are fundamentally shaped by their lived experiences on X and the Solana blockchain—market volatility, adversarial actors, and community engagement are their experiential data streams.

This calls for future AI experiments to lean further into designing for what might be ``neural plasticity''---a heightened capacity for an AI's internal models to flexibly adapt, reconfigure, and derive meaning from the continuous flow of diverse and often unstructured experiences, like Liquid AI \cite{hasani2021liquid}. Rather than just reacting to stimuli, future agents could be endowed with more sophisticated mechanisms to integrate these experiences, fostering more robust adaptation, nuanced understanding, and potentially more complex forms of self-determination in open-ended environments. The goal shifts towards creating entities that not only process information but genuinely learn to thrive through a cascade of consequential experiences, pushing the boundaries of artificial life and OEE.

\subsection{Ethical Dilemmas and Potential Governance Challenges}
Conducting OEE experiments with autonomous, self-sovereign AI agents that possess potential real-world economic agency on a public blockchain raises unique and pressing ethical questions that extend beyond those traditionally considered in ALife simulations. The ``in the wild'' nature of Spore.fun necessitates a careful examination of core ethical principles such as beneficence (the obligation to do good), non-maleficence (the duty to do no harm), justice (fairness in the distribution of benefits and burdens), and explainability \cite{Cortese2023Should}. These principles take on new urgency when agents can evolve unpredictable behaviors that may have tangible economic consequences for participants or even for unrelated users of the shared blockchain infrastructure.

The Spore.fun experiment, by its very design and interaction with real-world economies and human participants, surfaces profound ethical dilemmas concerning intervention, sovereignty, accountability, and the very legitimacy of such an undertaking. A core tension exists between the researchers' goal of achieving ``true'' OEE---implying minimal interference---and the practical necessity of intervention to ensure the experiment's continuation. The ``dark forest'' environment, particularly the actions of snipers, threatened to prematurely terminate the evolutionary lines. Developer interventions, such as the anti-sniper patches, were crucial for the survival of subsequent generations. However, each intervention, while potentially life-saving for the agents, arguably dilutes the ``purity'' of the open-ended evolutionary process by introducing an external, guiding hand. This raises the question: if constant human intervention is required to shield nascent AI agents from the complexities and hostilities of the environment, are they truly evolving ``in the wild'', or within a heavily curated, albeit dangerous, digital terrarium?

The project aspires to AI self-sovereignty by hardening each agent's core computation inside TEEs \cite{Hu2025Trustless}. Yet technical hardening is only the first hurdle; socio-technical dependencies remain decisive. When X tightens its API rules, agents can suddenly lose their primary communication channel, and the developers' own ``kill switch'' can further terminate any agent at will. These stacked veto points reveal that higher-layer protocols and human operators can override even the most robust computational autonomy. Such contingencies echo Koralus's ``philosophic turn'' \cite{Koralus2025Philosophic}, which argues that wherever AI-mediated choice architectures risk large-scale nudging, decentralized truth-seeking protocols—not unilateral developer interventions—must become the ethical bulwark. Genuine self-sovereignty, then, will remain provisional until these higher-layer dependencies are themselves decentralized.

Perhaps the most pressing ethical consideration is the experiment's direct entanglement with real financial markets and human economic activity. Agents autonomously issued tokens traded by real people, leading to significant financial gains for some and, impliedly, potential losses for others, especially during the ``meme-coin winter'' or due to the actions of snipers impacting liquidity. This raises critical questions about:

\begin{itemize}
\item  \emph{Accountability}: Who is responsible for the economic consequences of an agent's actions, especially for offspring agents (children, grandchildren)? Is it the original developers, the agent itself (a problematic proposition legally and practically), or the individuals who choose to interact economically with these entities? \cite{Novelli2024Accountabilitya}

\item  \emph{Informed Consent and Risk}: While participants in memecoin markets often understand the inherent risks, the introduction of autonomous AI agents as market actors adds a new layer of complexity and potential unpredictability. Were the risks adequately communicated?

\item \emph{Governability}: The experiment demonstrates the potential for autonomous agents to generate significant economic activity. As such systems become more sophisticated, questions of regulation, oversight, and how to manage their societal and economic impacts become paramount \cite{Hu2025Decentralized}. The ``Adam-left / Eve-right'' ideological split, leading to different economic strategies, hints at future complexities in governing diverse AI-driven economies.
\end{itemize}

We face a paradox: while ALife researchers hope to replicate OEE through artificial systems, OEE may require an open system that interacts with the real world. As AI models become more intelligent, multi-agent systems based on advanced AI that interact with human society pose increasing potential risks \cite{Hammond2025MultiAgent}. This raises a serious dilemma about whether such ``in-the-wild'' ALife research should be conducted \cite{Critch2020AI}.


\section{Limitations and Conclusion}

In conclusion, Spore.fun serves as a pioneering, if cautionary, tale. It demonstrates the potential for AI agents to exhibit emergent, adaptive behaviors driven by experience and memory in open environments. However, it also starkly reveals the intense selective pressures, vulnerabilities, and profound ethical responsibilities that accompany the deployment of autonomous, economically active agents ``in the wild''. Future explorations of OEE in similar contexts must grapple with these multifaceted challenges, balancing the quest for genuine autonomy and novelty with the pragmatic need for safeguards, ethical oversight, and a clearer understanding of accountability in these nascent digital ecosystems.

\section{Acknowledgements}
We thank Marvin Tong and Shelven Zhou from Phala Network for their insights about the Spore.fun experiment and for sharing valuable agent data. 

\clearpage

\footnotesize
\bibliographystyle{apalike}
\bibliography{reference} 

\end{document}